\documentclass[sigconf,authorversion]{acmart}

\AtBeginDocument{%
  }

\setcopyright{cc}
\copyrightyear{2025}
\acmYear{2025}
\acmISBN{}
\acmConference[CHI '25]{CHI Conference on Human Factors in Computing Systems}{April 26-May 1, 2025}{Yokohama, Japan}
\acmBooktitle{CHI Conference on Human Factors in Computing Systems (CHI '25), April 26-May 1, 2025, Yokohama, Japan}\acmDOI{10.1145/3706598.3713565}

\begin{document}

\title[Triangulating on Possible Futures: Conducting User Studies on Several Futures Instead of Only One]{Triangulating on Possible Futures: Conducting User Studies on Several Futures Instead of Only One} 

\author{Antti Salovaara}
\email{antti.salovaara@aalto.fi}
\orcid{0000-0001-7260-8670}
\affiliation{%
  \institution{Aalto University}
  \city{Espoo}
  \country{Finland}
}

\author{Leevi Vahvelainen}
\orcid{0009-0006-5314-6590}
\affiliation{%
  \institution{Aalto University}
  \city{Espoo}
  \country{Finland}
}
\email{leevi.vahvelainen@aalto.fi}

\renewcommand{\shortauthors}{Salovaara and Vahvelainen}

\begin{abstract}
Plausible findings about futures are inherently difficult to obtain as they require critical, well-informed speculations backed with data. HCI scholars tackle this challenge via user studies wherein futuristic prototypes and other props concretise possible futures for participants. By observing participants' actions, researchers then can `time travel' to see that future as reality, in action. However, such studies may yield particularised findings, inherent to study’s intricacies, and lack broader plausibility. This paper suggests that \emph{triangulation of possible futures} may help researchers disentangle particularities from more generalisable findings. We explored this approach by conducting a study on two alternative futures of AI-augmented knowledge work. Some findings emerged in both futures while others were particular to only one or the other. This approach enabled cross-checking of plausibility and simultaneously afforded deeper insight. The paper discusses how triangulating possible futures renders HCI studies more future-proof and provides means for reflective anticipation of possible futures.
\end{abstract}


\begin{teaserfigure}
  \vspace{0.5cm}
  \includegraphics[width=\textwidth]{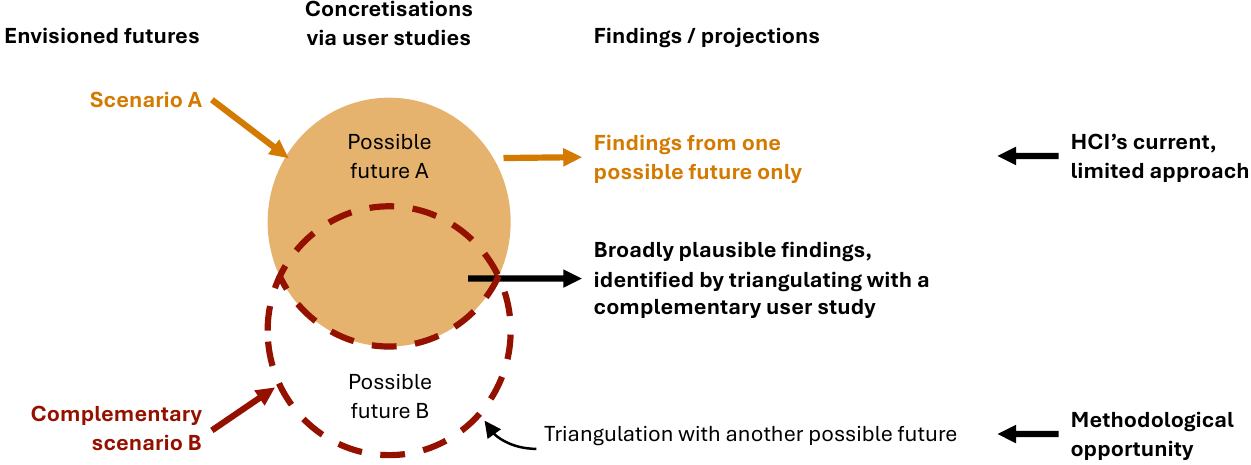}
  \caption{How investigation of a possible future benefits from triangulation via a complementary study.}\vspace{0.5cm} 
  \label{fig:money-shot}
  \Description{Figure 1: Venn diagram showing an overlap between two scenarios that are concretised as user studies. Some of the findings are common to both studies while others are specific to one or the other study. Figure suggests that if there is only one study, it is HCI's current, limited approach. Addition of another study is claimed to provide a methodological opportunity.} 
\end{teaserfigure}


\maketitle

\section{Introduction}

Nearly without exception, HCI studies are future-oriented. We envision and develop new technological augmentations; carry out user studies of emerging or recently introduced technologies, to learn about users' appropriation of them so as to inform design efforts down the line; and undertake critically oriented discussion of directions in which technology and society may be heading.
One of the main methods behind the field's future-oriented research involves organising a \emph{prototype-based user study}, in which some possible new technology is prototyped to such a level that it can be interacted with, after which evaluation of its features and underlying assumptions gets arranged. This entails recruiting individuals whose characteristics match those of the future users envisioned. Such studies sometimes go by the name `speculative enactments' \cite{elsden2017on,simeone2022immersive}, where the reference to speculation emphasises the work's future-oriented nature and `enactment' refers to staging and propping \cite{salovaara2017evaluation} of the study, designed to evoke a realistic user experience and prompt more authentic user behaviour. By means of its prototypes, props, and sometimes actors, this method lets researchers `time travel' to the future: albeit collected in the present, the empirical content allows us to draw inferences about the prototype’s future use -- i.e., to \emph{project} \cite{salovaara2017evaluation} `back to the future' wherein the technology might get used. This helps us anticipate future technology uses and create better designs \cite{moesgen2023designing}.

However, most HCI user studies confine themselves to single case studies. Each study investigates one scenario within a single possible future. This carries a risk, as limiting oneself to considering just one prototype and singling out a specific future can lead to poor anticipation of futures, distorted planning, and blindness to important uncertainties and perils. Because prototyping and staging realistic-feeling possible futures requires considerable time and other resources, we should optimise our efforts for the aim of obtaining findings with high \emph{plausibility}. That is, we should seek scenarios that are `believable and credible' even if it may be impossible to assign them probabilities or likelihoods (cf. \cite{fischer2021social,ramirez2014plausibility,schmidt-scheele2020plausibility}). 

Clearly, the plausibility issues arising from single case studies must be mitigated. In his paper, we present an exploration into \emph{triangulation} across several possible futures: investigating multiple possible futures within the research project. Figure~\ref{fig:money-shot} articulated benefits from triangulating possible futures. As the figure illustrates, even a single complementary study 
offers another look at the possible futures. With a two-future project, researchers may enact a pair of possible futures for user studies in such a way that the settings display suitable overlap yet differ in carefully considered respects. We posit that such two-pronged enactment should aid scholars in uncovering:

\begin{enumerate}
    \item \emph{Findings that are `broadly plausible', encapsulating characteristics across several possible futures:} Even if numerous other characteristics of the various futures differ and fluctuate, some general-applicability findings may be identifiable in many futures. Observing those characteristics in two complementary futures points to a greater chance that they are components of other futures too. 
    \item \emph{Findings that are `scenario-unique' by being characteristic of specific future scenarios only:} When only one enactment of possible futures produces a specific outcome (in conditions that show sufficient overlap), it can reveal something interesting that distinguishes that possible future in particular. Such a finding can assist in decision-making and planning, and in technology design that supports steering development paths toward preferable futures.
\end{enumerate}

This paper reports on exploring the benefits of this methodological principle in practice. We enacted two mutually distinct possible futures of AI-augmented freelance knowledge work in the context of investment-related analyses and decision processes. In one enactment, the AI agent’s role was to cultivate worker expertise and competence; in the other, the system exploited users' input to train an AI-based model, for eventual assimilation of their skills. 
Proceeding from the findings from both arms of the study, we reflect on the opportunities that triangulation grants to future-oriented HCI studies. We also discuss types of triangulation between possible futures and attempt to enrich HCI scholarship by supplying methodological recommendations that make the future of our field's research more future-savvy. 

\section{Background}
\label{sec:background}

Although triangulation is by no means a new concept in HCI methodology, its underpinnings in two academic fields – the social sciences' qualitative research and the psychology domain's experimental research – rarely get touched upon. Therefore, the subsections below present the conceptual basis for triangulation in empirical research. With this groundwork laid, we then cast a critical eye on the methods commonly employed in the HCI’s user studies, especially in work that involves prototypes and is decidedly future-oriented.

\subsection{Triangulation}

Triangulation has gained popular regard as a foundational element of rigorous empirical research. The main idea behind it is that applying mutually complementary research tools equips scholars to learn more about their research topic than reliance upon any of the components in isolation can offer. These elements should be chosen such that they offset one another’s weaknesses: each provides information that the others cannot offer. Thus, triangulation covers the ground by bridging gaps between approaches.

The term `methodological triangulation' was popularised already in 1959 by experimental psychologists \citet{campbell1959convergent}, in a quantitative context. They were interested in developing techniques whereby practitioners can distinguish among separate sources of variation in a measurement: undesired factors that stem from the measurement mechanism's weaknesses and aspects that reflect genuine variance in the object of interest alike. They suggested that using several methods enables identifying the measurement-internal weaknesses, and labelled it as `triangulation'.

Since then, the concept has been embraced more broadly, especially in qualitatively oriented social sciences. There, credit for its further development has often been accorded to \citet{denzin1970research}, who presented four ways of triangulating. Of these methods, presented in Table~\ref{tab:denzin-triangulations}, \emph{methodological triangulation} (Type 4) is the most commonly applied. It involves marrying multiple data-gathering and analysis techniques, such as observation and interviews, controlled and naturalistic research designs, and mixed methods. However, the table reminds us that engaging in triangulation might also rely on using multiple datasets (Type 1), recruiting several kinds of investigators to conduct the study (Type 2), and/or adopting several distinct theories as lenses for analysis of the data (Type 3).

\begin{table}[b]
    \caption{Four types of triangulation commonly employed in qualitative social-science research \cite{denzin1970research}} 
    \label{tab:denzin-triangulations}
    \begin{tabular}{c l p{5.4cm}}
        \toprule
        \multicolumn{2}{c}{Triangulation type} & Examples\\
        \midrule
       	1 & Data & Several participant groups, times, and/or contexts/locations\\
       	2 & Investigator & Analyses by researchers with complementary areas of expertise\\
       	3 & Theory & Analyses from multiple theoretical vantage points\\
       	4 & Methods & Lab vs. field, unobtrusive vs. interventionist, qualitative vs. quantitative, etc.\\
        \bottomrule
    \end{tabular}
\end{table}

Other researchers have returned to Denzin’s original idea. Among them are \citet{fusch2018denzins} and \citet{flick2018triangulation}, who  noted that triangulation not only affords \emph{better cross-validation} of findings but also offers \emph{deeper insight} into a phenomenon. Studies can also handle triangulation in a \emph{concurrent} and/or \emph{sequential} manner, distinguished by their temporal aspect -- in the former, all data-gathering for the phenomenon in question gets conducted at once, while the latter temporally separates the data-gathering components, as in the case of an observation stage followed by interviews \cite{creswell2003research}. Likewise, some triangulation follows a \emph{within-method} approach (e.g., a questionnaire may gather data from multiple viewpoints), while \emph{between-methods} work applies entirely different data-gathering procedures \cite{jick1979mixing}. Clearly, triangulation represents a rich conceptual landscape that furnishes researchers with a productive starting point for reflecting on, and improving, their methodological sophistication \cite{fielding1986linking}.

\subsection{Triangulation in HCI}

Most HCI literature on triangulation borrows from the social sciences and builds on Denzin’s typology. \citet{kaulio1998triangulation}, for example, have presented a model that adapts his categories to requirements engineering. It describes the methodological choices at play when differences appear in research locations, when investigators possess knowledge from separate areas, or when different methods are being used. Sensitised to the above-mentioned developments in the social sciences, \citeauthor{kaulio1998triangulation} have also discussed  between-methods vs. within-method triangulation, sequential and simultaneous stages' deployment, and an aim of validating hypotheses or deepening knowledge \cite{creswell2003research,jick1979mixing,fusch2018denzins,flick2018triangulation}.

HCI researchers have been particularly interested in capitalising on benefits from methodological triangulation. \citeauthor{pettersson2018bermuda}'s \cite{pettersson2018bermuda} meta-analysis of 100 UX studies found that 72\% had adopted triangulation in some form, with 46\% consisting of mixed-methods approaches (combinations of qualitative and quantitative methods). However, in marked contrast to that, only one UX research paper each had reported triangulating across theoretical frameworks (Type 3) or user groups (Type 1). HCI's interest in methodological triangulation is visible also in the debate about the benefits of lab vs. field studies \cite{kjeldskov2004is,kjeldskov2014was,rogers2007why}. However, these exchanges of opinions have not oriented themselves to the possibility of the methods mutually strengthening each other. Instead, the parties have considered methods in isolation, merely pointing to the weaknesses in one form of research and the strengths in another.

Although empirical studies applying theory or data triangulation, in turn, have been scarce, more conceptual HCI methodology papers have acknowledged the opportunities that they represent. In a spirit similar to social science \cite{fusch2018denzins,flick2018triangulation}, \citet{turner2009triangulation} have noted that triangulation can be either \emph{soft}, confirming knowledge by deepening it, or \emph{hard}, permitting hypotheses or assumptions to be falsified. They exemplify these approaches by drawing from gaming presence research and by applying a  phenomenological framework to it. \citet{mackay1997hci}, meanwhile, have drawn attention to HCI's role as a bridge between theories and observations: models of interaction, on one hand, and their empirical evaluations, on the other. They have suggested that embracing this standpoint allows for designs that are triangulated from these two, complementary directions. Also \citeauthor{van-turnhout2014design} \cite{van-turnhout2014design} have identified HCI as an interdisciplinary field that informs and cross-validates knowledge and designs from diverse angles: fieldwork, literature, workshops (i.e., prototyping), lab studies, and showrooms (submitting one's creation to others’ critique).

Finally, the ethos of triangulation is echoed in HCI's design literature, where scholars note that prototyping allows for triangulation. This happens when several design ideas are considered, each one presenting a distinct approach to satisfying multifaceted requirement sets. When a designer constructs prototypes of different kinds, the creations manifest different kinds of `filters', through which the design problem is approached from multiple angles \cite{lim2008anatomy,halskov2021filtering}.

\subsection{Triangulation in HCI's Future-Oriented Methodology Papers}

Regarding HCI papers that focus on probing possible futures and technology uses in them, we are aware of only one that mentions triangulation: \citeauthor{mankoff2013looking}'s \cite{mankoff2013looking} adaptation of the Delphi-based scenario method. Even there, triangulation is referred to only fleetingly, without elaboration. Still, various design methods for HCI that generate mutually distinct future-oriented outcomes have applied triangulation at least implicitly. Among those methods are \emph{speed dating} (for rapid scenario development \cite{davidoff2007rapidly}) and \emph{paratyping} (a prototyping-based research approach for studies on situated experiences \cite{abowd2005prototypes}). When these methods deal with contrasting viewpoints and involve comparing the viewpoints with each other, so as to validate or deepen the future-oriented insight, they make use of triangulation. This is true also of the so-called \emph{2x2 matrix method} employed in futures studies for creating contrasting scenarios \cite{vantklooster2006practising}. 

\begin{figure*}[t]
  \includegraphics[width=\textwidth]{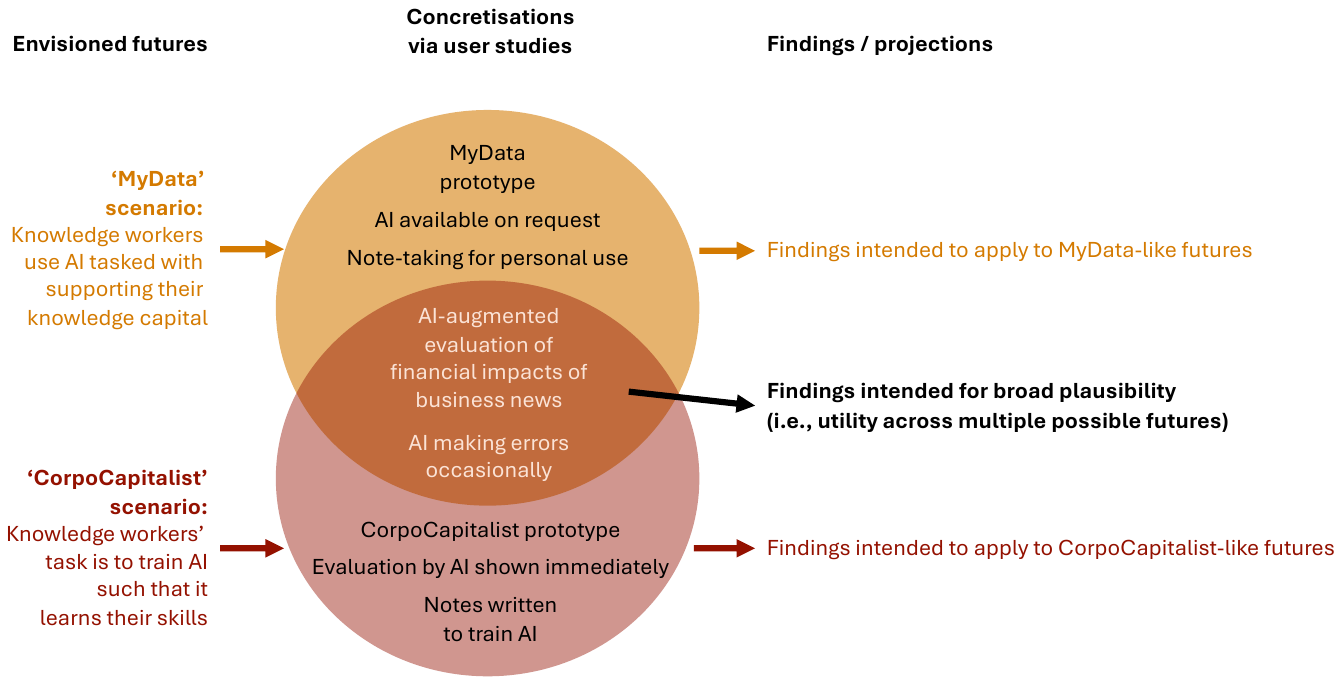}
  \caption{Triangulation of possible futures in our user study.}
  \label{fig:mydata-corpo-triangulation}
  \Description{Venn diagram that illustrates the similarities and differences between two studies called "MyData" and "CorpoCapitalist". MyData study's prototype offered AI that was available on request, and note-taking for personal use. CorpoCapitalist study's prototype showed AI's evaluation without asking, notes were asked in order to retrain AI. Common characteristics were AI-augmented evaluation of business news and that AI makes errors occasionally.}
\end{figure*}

The paucity of discussion specific to triangulation for future-oriented studies, notwithstanding the topic's prominence in HCI methods generally, might come as surprise. However, discussions on triangulation are often connected with validity, a concept linked to empirical data that exist in the present. Accordingly, it does not fully dovetail with phenomena that may take place in the future. In fact, some futures studies and futures foresight scholars (e.g., \citet{bell2003foundations-vol1}) maintain that research into futures is always speculative, whereby ascertaining validity of such findings is a conceptual impossibility. 
In consequence, futures studies deems it more important that foresight methods \emph{sensitise} the interested parties to possible future outcomes. Because triangulation's relevance often is most evident in the realm of validity, the concept is rarely discussed in futures studies.

Since information about a future cannot be validated in the present, the futures studies field can be said to generate not \emph{knowledge} (`justified true beliefs' \cite{chisholm1989theory}) but socially constructed \emph{representations of the future} \cite{clardy2022what}. While the representations may nevertheless encompass descriptions of some shared objective reality, those descriptions are constructivist understandings of the reality. Hence, futures studies employs the term `credibility' in place of `validity' in its criteria for high-quality research \cite{clardy2022what}.

By corollary -- because HCI's future-oriented studies can be considered an empirical variant of futures studies -- the foregoing logic extends to HCI. This may explain why HCI's future-oriented methods, whether attending to prototype-based studies \cite{salovaara2017evaluation}, VR-realm simulations of possible futures \cite{makela2020virtual,simeone2022immersive}, or speculative enactments \cite{elsden2017on}, do not mention triangulation. Likewise, papers focused on means of envisioning futures (such as workshop-anchored brainstorming \cite{epp2022reinventing} or bodystorming \cite{oulasvirta2003understanding}) or on more speculative methods (e.g., design fiction \cite{lindley2017implications,coulton2017design,pargman2017unsustainability,kozubaev2020expanding}) are silent on the matter. 
We can conclude that the possibility of triangulating possible futures has not been considered before in HCI literature and warrants further examination.

\section{A Study of Triangulation across Possible AI Futures}

To explore the triangulation opportunities offered by studies on possible futures, we organised an exploratory study with two possible futures of AI-augmented knowledge work.
We explored how the patterns of findings might overlap if the futures' conceptual underpinnings had nearly opposing 
worldviews. 
The motive for choosing these futures lay in ongoing debate, in both academia and the popular press, about augmentation vs. automation via AI. 
Figure~\ref{fig:mydata-corpo-triangulation} depicts our outline of the two possible futures' triangulation with each other.
We considered that overlapping findings could spotlight broadly plausible projections about futures while differences between the two enactments should reveal intriguing contrasts that, though not promising broad applicability, would shed light on the individual futures separately.

\begin{figure*}[t!]
  \includegraphics[width=\textwidth]{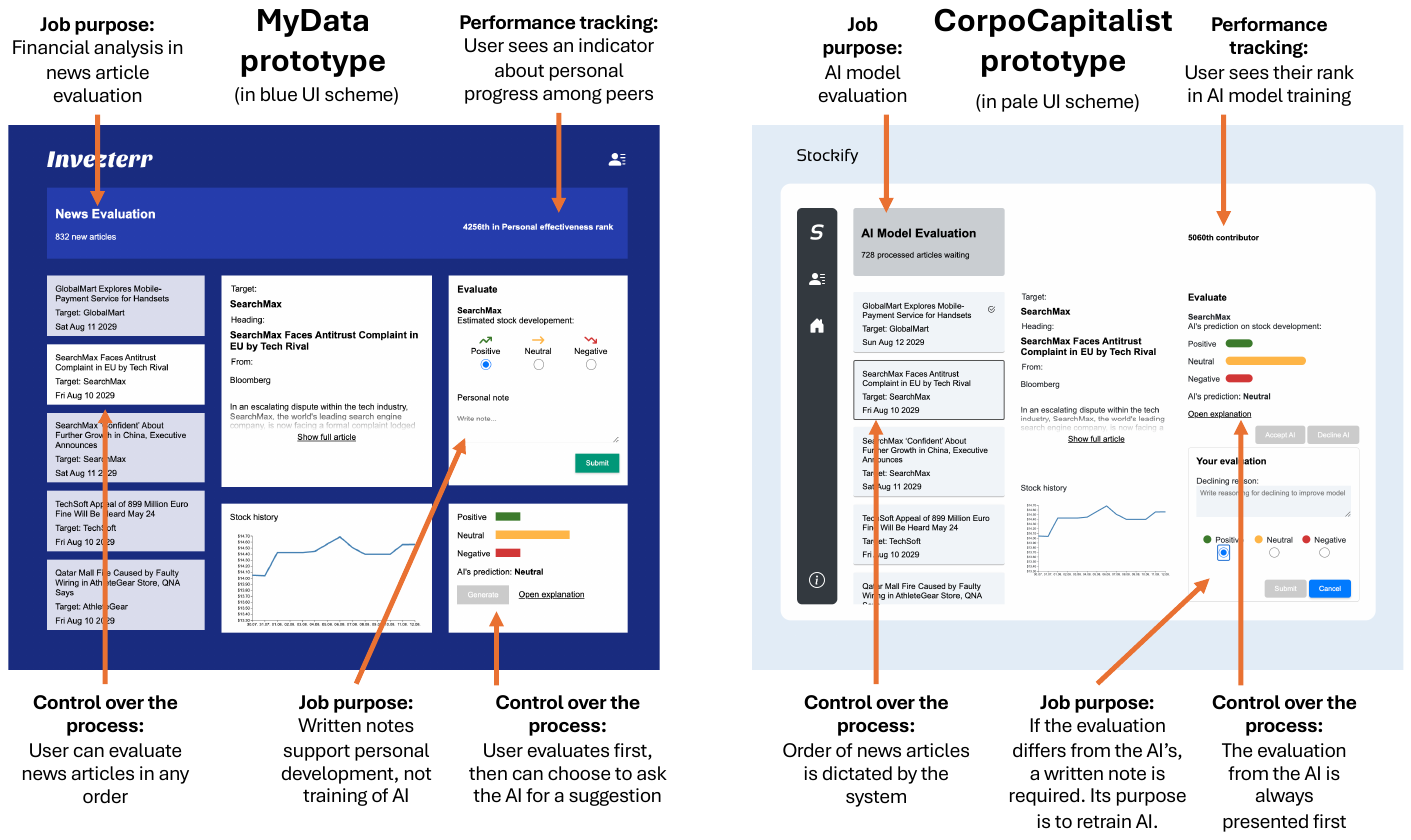}
  \caption{Differences in the main screen between the prototypes.}
  \label{fig:UI-differences}
  \Description{Screenshots of MyData and CorpoCapitalist UIs showing how the MyData prototype provided more control to the user. MyData prototype's subtitle was "News evaluation", its performance tracking informed about personal progress among peers, its user could choose the order by which they evaluate news and when they see AI's suggestion, and the note-taking's purpose was said to be related to self-improvement. CorpoCapitalist prototype's subtitle was "AI model evaluation", its performance tracking informed about user's contribution to AI's re-training, the user did not have control on the order of evaluation tasks nor when AI's recommendation is shown, and the note-taking purpose was communicated as a re-training information provision.}
\end{figure*}

We characterise the two AI-linked futures as follows:
\begin{itemize}
    \item \emph{The `MyData' future}: Inspired by activism for protecting the ownership of one's personal data \cite{poikela2020mydata}, the AI in this future lets workers exercise extensive control of their AI tools. It supports accrual of the worker’s knowledge capital by acting as an agent able to supply second opinions, counterarguments, and contrasting viewpoints. The worker controls the interaction with the AI agent, and the data are generated to support professional growth and self-improvement. All this work by the AI is directed also toward growth in companies’ knowledge capital.
    \item \emph{The `CorpoCapitalist' future}: Inspired by recent widespread fears of dystopian mass unemployment due to AI’s burgeoning intellectual capacity, this future displays elements of corporate capitalism wherein employees have little control over their work and the purpose is to support corporate objectives of profit maximisation. The AI agent is not under employee control. Instead of supporting workers, it acts to support companies’ efficiency goals by drawing on knowledge workers’ expertise as a source to train continuously improving machine-learning models. Worker interactions with the AI therefore operate as a mechanism to translate human skills into a format that leads toward automated decision-making.
\end{itemize}

These names for the two futures -- MyData and CorpoCapitalist --  are only used in this paper. In user studies, the participants interacted with two futuristic prototypes called Invezterr and Stockify, respectively, and we used those names also in the interviews when probing about the experiences that they had had.

In addition to developing plausible scenarios for futures from the starting points described above, we sharpened the futures' concretisation by injecting two specific AI-linked elements into them. Firstly, we were interested in the phenomenon of \emph{latent deskilling}, which poses a threat bound up with rising automation. Automation makes humans liable to outsource their decision-making to AI and grow complacent, overlooking its defects. Over time, `skill erosion' may develop \cite{rinta-kahila2023vicious}. The second component factored in is AI's potential to err radically in the face of even the tiniest differences in its input data, a problem sometimes described as a `jagged frontier' in the boundary of what the AI may master \cite{dellacqua2023navigating}. One of the jagged frontier's corollaries is that AI holds utility for high-reliability tasks only if combined with human monitoring \cite{salovaara2019high}. 

\begin{figure*}[t]
  \includegraphics[width=\textwidth]{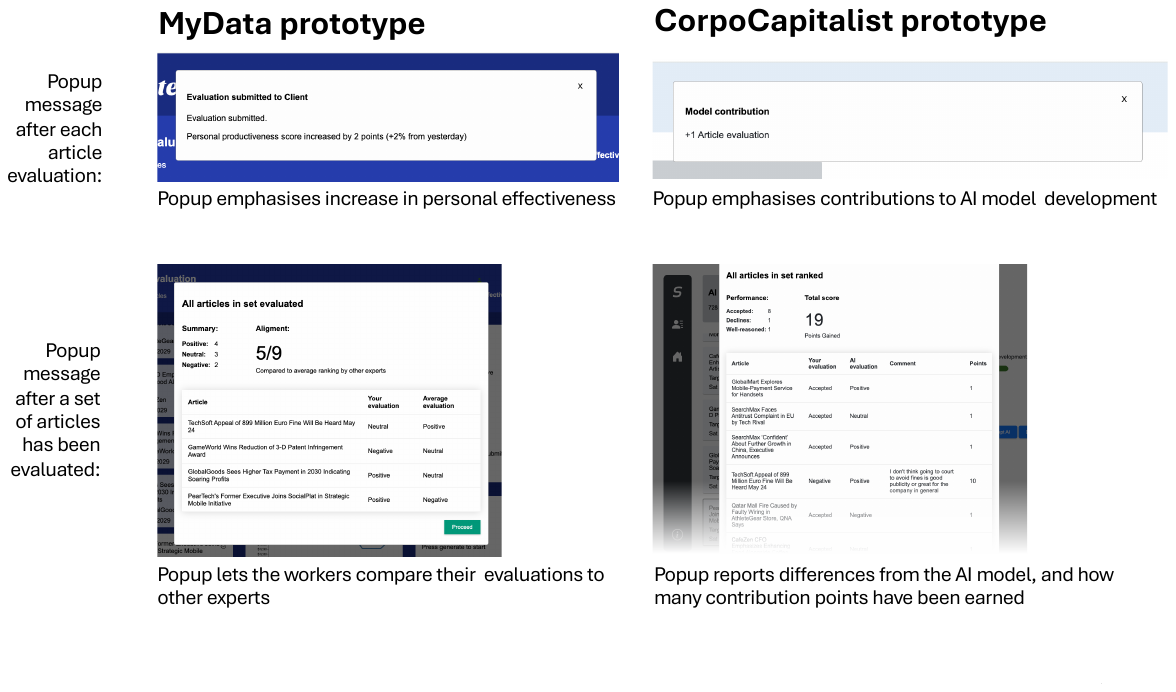}
  \caption{How the prototypes' evaluation confirmation popups differed.}
  \label{fig:UI-summary-differences}
  \Description{Screenshots of MyData and CorpoCapitalist UIs' summary popups. Popup after every evaluation highlights personal effectiveness in MyData but contribution to AI training in CorpoCapitalist. Popup after a set of evaluations highlights comparison to experts in MyData but to AI vs. worker differences in the CorpoCapitalist prototype.}
\end{figure*}

From these starting points, we developed an overarching scenario of business news analysis directed to informing financial investments.
We asked participants to imagine being freelance workers who monitor business-related newswire articles from agencies such as Bloomberg and Thomson Reuters. As a freelancer, each participant would work for one company in the morning and another in the afternoon. In our research design, one of the companies, named `Invezterr', corresponded to the prototype of the same name (aligned with the MyData future) and the other (`Stockify') represented the CorpoCapitalist one. Figures~\ref{fig:UI-differences} and ~\ref{fig:UI-summary-differences} present both prototypes -- our main instruments for enacting the two futures. With emphasis on their differences and shared characteristics both, the diagrams capture the MyData prototype's objective of accentuating learning opportunities within a community of peers alongside the CorpoCapitalist one's highlighting of the worker's contribution to AI-model development and performance of that role.

The participant's task was to read short news items and assess whether they imply positive or negative stock price impacts for the companies mentioned. Our development of this work scenario took inspiration from research that has applied sentiment analyses to such datasets and demonstrated that news briefs can predict stock price changes \cite{upreti2019knowledge-driven}.

Our scenario wherein each work day comprises interaction with both prototypes, separately (one in the morning, the other in the afternoon), supported creating a within-subjects research design that let us study each participant's experiences and individuals' reactions to both futures. This also allowed us to interview the participants so that they could compare and reflect on their experiences.

As Figures~\ref{fig:UI-differences} and \ref{fig:UI-summary-differences} attest, the two UIs 
shared many similarities. Both presented a sidebar containing a set of 18 business news articles that the participant was to evaluate -- all authentic news articles run in recent years, though modified such that the company names had been replaced with fictitious ones. The UI also displayed corresponding charts accurately reflecting stock prices, again without revealing the company’s true name.

\begin{figure*}
  \includegraphics[width=\textwidth]{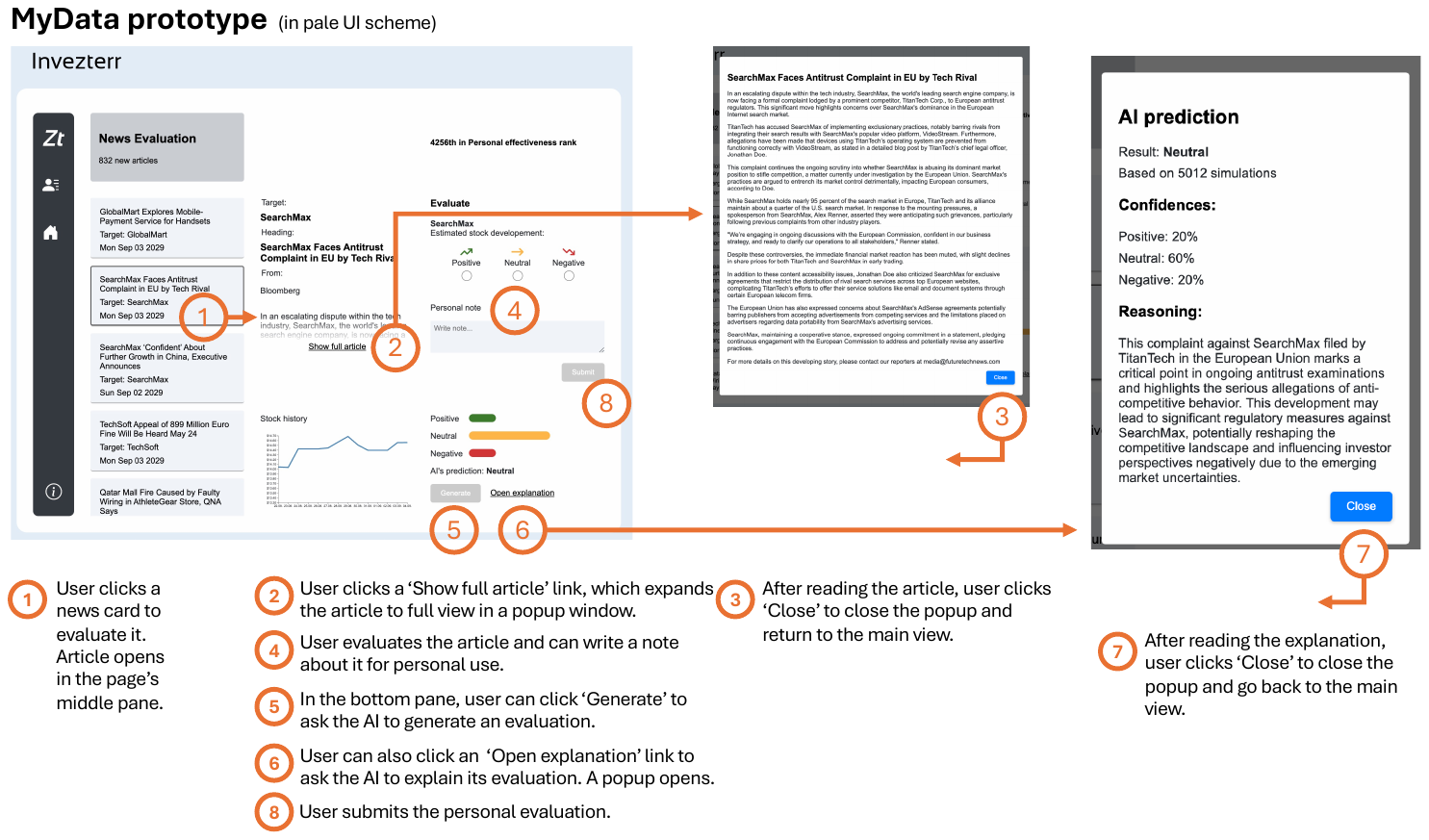}
  \caption{The interactions involved in evaluating a news article, in the MyData condition in its UI's pale-colours version.} 
  \label{fig:UI-flow}
  \Description{Three screenshots from the UI that describe through eight steps a possible interaction sequence in both prototype's use: 1. selection of article, 2. opening it to full view, 3. reading the an article, 4. closing the full view, 5. commanding AI to provide suggestion, 6. opening AI's verbal explanation to its suggestion, 7. closing the explanation, 8. submitting user's evaluation.}
\end{figure*}

Figure~\ref{fig:UI-flow}, portraying MyData (Invezterr), presents the interaction flow, which was identical between prototypes. 
The participant’s task was to evaluate each news article as having a positive, neutral, or negative impact on the stock price of a given company mentioned in the article.
An AI agent would provide its own evaluation, accompanied by a textual explanation for said conclusion. 
The AI agent was a fake facade, however, designed into the UI prototype to give the participants a sufficiently authentic impression of working with AI.

We pre-generated all the evaluation recommendations that the AI would offer. Thus also every participant saw the same recommendations. 
For 15 of the 18 articles shown, we prepared `sensible' recommendations: consistent with the true sentiment of the article content, the AI suggested a positive, neutral, or negative stock-price impact. For the remaining articles, we prepared an erroneous conclusion, to capture cases wherein AI performance might unexpectedly falter -- reflecting the possibility of the above-mentioned jagged frontier.
For all evaluations, we also prepared textual explanations that the pseudo-AI would display to the participant. These were short paragraphs of reasoning that we made `AI-like' by means of ChatGPT 4.0: Prompting it with the news article and direction of impact as input, we requested a brief justification as the output.

Finally, to convey a sense of users' performance being monitored, the UI rewarded participants in several ways. For example, when the user diverged from the AI's prediction and explained the rationale behind it, or provided other written notes about the decision, the work-performance ranking 
shown by the UI rose (see Figure~\ref{fig:UI-differences}). Also, at nine-article intervals, we supplied summaries of the participant's decisions in absolute terms and either relative to peers (in the MyData condition) or as a score for contributions to training the AI model (in the CorpoCapitalist setting).

Each participant completed similar tasks with both UIs. We counterbalanced the order with regard to 1) the first prototype presented (MyData vs. CorpoCapitalist), 2) the colour scheme chosen for each (a blue or a pale UI, which could be used interchangeably without any effect on either prototype's interactions or on UI-element locations), and 3) which of the two pre-created sets of news articles was used for each.

\subsection{Research Design}
\label{sec:research-design}

The goal for our study was to demonstrate how possible futures can be triangulated.
Because HCI's future-oriented findings are representations of possible future(s), not factual knowledge, the methodology exhibited a constructivist undercurrent. That is, while the study involved empirical data, our interpretations were actively constructed conjectures about states of affairs with potential to arise in the future (e.g., \cite{fuller2009constructing, dejouvenel1967art}). Even if taking a constructivist tack, we strove to learn about possible futures via mechanisms in ways that are not relativistic but could be considered credible 
by many researchers instead. 

Our interpretations were rooted in both qualitative data (especially from observations and interviews) and quantitative material (questionnaire responses) \cite{olsen2004triangulation}. Albeit constructivist, our design followed two methods that are commonplace in realist approaches (i.e., in scholarship focused on identifying causal mechanisms that could explain the events examined or patterns seen in the data \cite{sayer1999realism}) -- namely, comparative analysis, on which triangulation relies, and interventions, introduced by our prototype-based study design. Neither method is exclusively realist, however: deliberate comparisons are essential also in qualitative research (e.g., in grounded theory's `constant comparative method' \cite{glaser1967discovery}), and interventions form the basis for action research \cite{mcniff2013action}. 

The study employed a within-subjects design wherein every participant experienced engagement in both futures through the task of analysing 18 news articles in interaction with each prototype (one set of 18 unique items per prototype).
Following both the national and our university’s own ethical guidelines, the study did not require an inspection and approval from an IRB.
We asked the participant to imagine being a freelancer whose daily work consists of a morning session in one company and an afternoon session in another. To instil this mindset among the participants, we presented the following scenario: 

\begin{quote}
	In a future 5--10 years away, two companies -- Invezterr and Stockify -- are offering investment
advice partly based on media discussion. Both companies rely heavily on the use of AI technologies in forming holistic evaluations. However, these models are imperfect and occasionally fail to catch surprising clues from written information. Thus, both organizations need human workers to provide high-quality insights. How they use the workers is what sets the organizations apart: while one of them sees their workers as a valuable asset that are supported through AI technology, the other sees the value in their AI and uses human workers as a necessary resource in correcting the occasional mistakes until the mistakes can be fully eliminated. Both of the organizations use the same technology provider that has tailored the system according to the needs of each organization. 
\end{quote}

After introduction of this scenario came the bulk of the experiment, split into five parts. Part 1 represented the participant's morning, devoted to work with one of the companies (chosen randomly in advance). Here, the user evaluated a set of 18 news articles. 

In part 2, each participant completed a short questionnaire and answered a few interview questions: Immediately after the experiences of using the prototype, we used seven-point Likert items, adopted from studies on technology acceptance and job satisfaction \cite{venkatesh2003user,moore1991development,morgeson2006work}, to enquire about the participant's attitude to utilising the technology in question (via the `Attitude toward using technology' scale \cite{venkatesh2003user}), how well it meshed with the participant's preferred work practices (via the `Compatibility' scale \cite{venkatesh2003user,moore1991development}), and the task's perceived value (or `Task significance' \cite{morgeson2006work}). The user was asked to reflect aloud on the reasoning behind the responses while completing the questionnaire. We finished the second part by interviewing the participant briefly about the coping mechanisms related to the changes that the scenario envisioned would entail for working life. The interview protocol was informed by the Coping Model of User Adaptation (CMUA) \cite{beaudry2005understanding,elie-dit-cosaque2011opening}, which suggests that people formulate their way to deal with IT-related changes by 1) assessing the balance of whether the change poses a threat or presents an opportunity, then 2) considering their degree of control over ensuing events (low/high). 
CMUA posits four possible coping mechanisms: benefits maximising (high-control opportunity), benefits satisficing (low-control opportunity), disturbance handling (high-control threat) and self-preservation (low-control threat). We had one question about threat assessment, control assessment and coping mechanism adoption each (see Appendix~\ref{sec:post-scenario-interview}). 
After this, a short `lunch' break marked the end to the metaphorical morning.

Parts 3 and 4 followed, analogous to the previous two parts. Having spent the `afternoon' with a different company, the participant then went through a final interview: in part 5 of the experiment, we asked about experiences of both halves of the notional work day (see Appendix~\ref{sec:concluding-interview}). Enquiries focused on such matters as how the users would express the differences between the UIs, what they would think about these kinds of AI-augmented financial analysis jobs, and whether they had noticed anything exceptional in the evaluations that AI had provided. At the very end of the interview, we solicited broader reflections on AI’s implications for work from social, technological, economic, ecological, political, legal, and ethics angles (i.e., within the STEEPLE framework \cite{schwartz1991art}).

Although we expected user satisfaction, congruence with preferred work practices, and the task’s perceived significance to be greater in the MyData future, and that the answers about coping would reveal higher control ratings and a sense of opportunity more than threat also in the MyData future, we did not formalise these assumptions as strong hypotheses; submitting them to truth-testing would have been at odds with the idea that future-oriented studies investigate possible future states-of-affair instead of objective phenomena.

\subsection{Analysis}

Since the sessions incorporated two interactive parts and three occasions for taking stock via interviews, our dataset was qualitatively rich. It seemed fitting, then, to concentrate on qualitative analyses, angling our study for abductive findings. We sought patterns in the differences between how participants experienced the CorpoCapitalist and the MyData future and in their associated reflections. Once we were armed with empirical findings, we could embark on methodological reflections, considering ways in which those findings could inform scholarly use of triangulation for studying possible futures. 

Because the participants evaluated their experiences with numerical ratings after both the `morning' and the `afternoon' session, we were equipped to carry out quantitative analyses too. Using repeated-measures ANOVA, we analysed the differences in the participants' questionnaire responses with regard to use attitudes, compatibility with one's preferred practices, and task's perceived significance (again, see Appendix~\ref{sec:post-scenario-interview}).
To compare the coping mechanisms that participants linked to the CorpoCapitalist and MyData future, we performed McNemar's test \cite{agresti2002categorical}, which is a within-subjects version of Chi-squared tests for comparisons between counts for categorical variables: whether the future presented a threat or an opportunity and whether the respondent sensed control over the situation vs. a lack thereof.

\section{The Findings}

We recruited 13 members of our university's entrepreneurship- and business-related interest groups via social media. All had experience of business and/or economics, mostly from their studies but some also via entrepreneurship or other employment. Their ages ranged from 10 to 31 (mean: 25.5). In free-text responses, three identified as female, one as a woman, and nine as male. As for other background, six characterised themselves as Finnish while the others were from Brazil, China, Hungary, India, Syria, and Turkey.
All sessions with participants were completed successfully; therefore, there is no need to account for missing data. 
Uniform conditions supported further robustness, thanks to the paper's second author conducting all of the sessions.

We believe that we enacted reasonably believable futures with the prototypes. Our staging of the possible futures gave the participants room for successful imagining. Six of them did, however, comment on the characteristics of the task: evaluating financial impacts on the basis of isolated news articles did not feel entirely plausible. These participants articulated a desire for opportunities to seek and obtain more information, beyond the articles. Also, two participants commented that their business school education had taught them that, in general, it is practically futile to try to predict stock market impacts and `beat' the average market index in the long run. Despite these misgivings, all participants felt able to relate to the tasks they performed and also admitted that, regardless of their doubts, they could readily believe that some firms may start carrying out AI-assisted evaluations of how news articles affect stock markets. 

In general terms, our analysis endeavour was centred on finding similarities and differences between the futures the two prototypes represented. We were interested especially in the participants' experiences of these possible futures and in their reactions to the futures. We begin by reporting on the quantitative analyses. The we build on that foundation by painting a picture of the qualitative findings.

In non-statistical terms, the analyses (which we report in more detail below) found out that participants' system use attitudes for MyData prototype were more positive than for the CorpoCapitalist prototype. We did not find differences from the other constructs, however. Although a pure average suggested that participants agreed with AI's suggestions more often in the MyData prototype, this difference was not statistically significant. The biggest differences appeared in coping mechanisms: the future portrayed by the MyData prototype was much less threatening than the one by the CorpoCapitalist prototype.


Let us now elaborate on the analyses summarised above. For each construct, 
we created an aggregate variable by averaging the responses for the respective items. Though we used validated scales, the thresholds for the constructs' reliability and validity were not fully met: the Cronbach's alphas of 0.78, 0.56, and 0.82 for system-use attitudes, compatibility with one's work approach, and task significance, respectively, suggest in particular that the compatibility construct was not reliable. The factor loadings using varimax rotation imply that, while task significance items loaded on their own factor as predicted, the items for attitude and compatibility clustered as two factors in a mixed order, differently than predicted. Nonetheless, our reporting on the analyses is based on the theoretically predicted item groupings. 

All of the variables passed the Shapiro--Wilk normality test. 
For the MyData prototype, the aggregated variables' averages were 5.52, 5.03, and 4.13 (on the 1--7 scale) while the corresponding averages for the CorpoCapitalist prototype were 4.79, 4.56, and 4.03.
According to within-subjects ANOVA tests, the averages for system use attitude differed significantly between the two prototypes (5.52 for MyData vs. 4.79 for CorpoCapitalist, $p = .01$, $\eta^2 = .44$), suggesting a large effect size ($ \eta^2 > .25$; \cite{cohen1988statistical,tabachnick2007using}). 
The differences for the compatibility with the preferred work method (5.03 vs. 4.56, $p = .27$, $\eta^2 = .10$) and for the task's perceived significance (4.13 vs. 4.03, $p = .68$, $\eta^2 = .02$) were not significant.

Turning to the participants' business news analyses and their agreement with the evaluations presented by the AI agent, we calculated four separate variables: averages of participant-level percentages when they agreed with AI's sensible recommendations and non-sensible agreements, separately for MyData and CorpoCapitalist prototypes.
Shapiro--Wilk tests revealed that agreements with the MyData AI's non-sensible suggestions were not normally distributed. Neither were user agreements with the CorpoCapitalist prototype's sensible suggestions. For both, we therefore used Friedman's test instead of repeated-measures ANOVA.
Although, on average, participants agreed less often with AI's suggestions in the MyData condition (69\% in cases wherein the evaluations were sensible and 56\% when they were not) than in the CorpoCapitalist condition (with corresponding values of 75\% and 59\%), none of these differences showed statistical significance.

Finally, the users' answers about coping mechanisms differed sharply between the two possible futures.
All but one of the 13 participants considered working with an AI companion as an opportunity, not a threat -- in both futures. In the subsequent assessment of whether they felt a sense of control over this opportunity, nine of them identified a high level of control over the situation in the MyData future while only four did so when contemplating the CorpoCapitalist future. 
According to McNemar's test, this represents a statistically significant difference in the perceived amounts of control ($p = .03$).
In the CMUA model's nomenclature, these results are consistent with perceiving MyData use as benefit-maximising while seeing CorpoCapitalist's use as benefits-satisficing. Participants' interview answers, analysed next, echo this pattern.

\subsection{Interaction Patterns}

In this subsection we will describe how the users interacted with the prototypes. Against that backdrop, the following subsection delves into how they experienced the futures. This discussion provides the basis for the third subsection's analysis of how successfully the study triangulated possible futures -- one being the MyData future and the other one the CorpoCapitalist future.

We witnessed a whole host of patterns in the participants' completion of their evaluations. Partly because the CorpoCapitalist prototype restricted interaction options (e.g., in that the system-generated evaluation was always available, with no opting out possible, and could never be suppressed), the participants developed several workarounds to circumvent disturbances to their work. Firstly, many of them stated that they tried to avoid looking at the AI's evaluation and thereby absorbing biases in the course of formulating their own opinion. Others, in contrast, read the evaluation justification because, in some respects, it provided a concise summary of the news article. This helped them absorb the details of the task more quickly, then formulate their own evaluation. A third group of participants adopted an approach of interleaved reading: they judged the news article alongside the rationale by iterating between the two -- going `back and forth' to identify differences between their interpretations and the ones offered. Finally, some participants had no worries about ending up biased after reading the system's evaluations. They claimed this to be because they regarded the assessments as \emph{a priori} erroneous so analysed them with a critical eye. They viewed their own evaluations as input to retraining the AI, whereby they would serve as content that could help eliminate AI's weaknesses.

The MyData prototype, on the other hand, saw much less heterogeneous use. It was almost uniformly used in one way alone: the participant chose one of the news articles for analysis, investigated it, occasionally wrote notes for personal reference in the text box provided, and selected their assessment using the radio button set. They then  clicked the button in the bottom of the screen and generated an evaluation by the model, and could then compare their own assessment to that of the model. Finally they clicked the Submit button. 

\subsection{Experiences}

Participants' experiences reflected their interaction patterns. At the most basic level, both prototypes sparked positive reactions to 
AI-augmented work overall. These positive reaction aside, the two prototypes led to quite easily distinguishable experiences, with the MyData prototype often being favoured over the CorpoCapitalist one.

\subsubsection{AI as a Knowledge Work Companion.}

As noted above, participants' recognition of the CorpoCapitalist prototype's utility was centred on its succinct summaries of the news content and its ability to explain how a certain evaluation could result. In the post-use debriefing, a few participants did mention detecting that the AI's evaluations appeared to include errors. It seems reasonable to assume that the three bogus evaluations deliberately included in both article sets were responsible for this belief and for these participants having grown 
gradually more suspicious of the `generated' evaluations as the study progressed. Still, the erroneous analyses did not offset the system's perceived utility for them, and most participants failed to spot mistakes in the evaluations -- or at least did not deem them noteworthy. 

As for the MyData prototype, the already earlier mentioned preference for it was nearly universal. Users appreciated that, unlike the CorpoCapitalist one, it let them read the news articles and arrive at opinions of their own before requesting a second opinion from the AI. For instance, when addressing the questionnaire item `The system makes work more interesting', one participant stated:

\begin{quote}
I definitely strongly agree with that. Yeah. And that's because [\ldots] it's something that you can always compare [your conclusion] to, and you can always identify to see whether your ideas go with what the AI model said. (P11)
\end{quote} 

The participants felt that the MyData prototype let them validate their analyses and provided an opportunity to improve their skills. This facet of the experience was accentuated by the possibility to write personal notes alongside the positive/neutral/negative evaluation.

\subsubsection{Performance Tracking Gamified the CorpoCapitalist Prototype.} 

The CorpoCapitalist prototype's unique effect was that it enacted perceptions of gamification.
The most important factor in this perception was the design of its performance-tracking feature (visible at the upper right in Figure~\ref{fig:UI-differences} and the feedback mechanisms shown in Figure~\ref{fig:UI-summary-differences}). A user who disagreed with the suggested evaluation and entered some justification in the text box was rewarded with a boost to the ranking 
as a contributor to the model. For some users, this was a highly motivating feature of the prototype:

\begin{quote}
It was fun [\ldots]. Because it was fun to see if you have the same opinion as or a different opinion from that AI [\ldots], and then, also, when you had to explain your opinion, that made it more interesting. (P8)
\end{quote}
\begin{quote}
I feel like, uh, I might have maybe put more effort into Stockify [the CorpoCapitalist prototype], because you also got more points. The points benefit the system, so [you were helping] whenever you [rejected the AI's offering and wrote an explanation] [\ldots]. That's why I feel as if I was more critical because whenever I saw there was something wrong, or if the model was wrong, then I would get more points for that. So, in that sense it was more rewarding to be more critical. (P11) 
\end{quote}

On the other hand, the participants reckoned that gamification rendered the CorpoCapitalist experience more effortful, since the goal of training the AI demanded relatively formal phrasing of the justifications. Because the explanations in the MyData condition were optional and intended for later personal reference, they offered flexibility for free-form expression of one's thoughts and thus less effort.

\subsubsection{Model Contributor Role Led to Reflections on Work's Wider Impact.} 

Another pattern emerged only in the CorpoCapitalist setting. This too was related to its emphasis on the worker's task of training an AI model. Participants were keenly aware that their corrective actions could improve AI performance. Those actions' impact could scale massively, thanks to the AI's potential for evaluation of numerous news articles. Some participants found this a strongly motivating aspect of the prototype:

\begin{quote}
	I guess I feel as if I'm helping it make better choices in the future. (P10)
\end{quote}

\begin{quote}
	[Upon presentation of the statement `The results of my work are likely to affect lots of other people'] I definitely agree with that. As I'm telling the model that something is right or wrong, it definitely influences other people that use the model. Yeah, so it has a really big impact [\ldots]. When you're training the model, you're making sure it reacts in a certain way and of how it is going to use that in the future. Can either use it correctly or incorrectly, so it has a really big way [of influencing] how other people use it. (P11)
\end{quote}

\begin{quote}
	[In consideration of the statement `The results of my work are likely to affect lots of other people'] Well, that's hard to predict, but I'll maybe answer `agree', because if that system becomes good and then it's being used somewhere\ldots Then probably it will then have impact. (P8)
\end{quote}

We will delve further into the training work's wider positive impact when reflecting on the benefits of triangulation.

\subsubsection{Peer Comparisons Foregrounded Learning in the MyData Prototype.}

While gamification and scaling of one's contributions were identified uniquely in the CorpoCapitalist future, the MyData prototype manifested unique outcomes of its own: it oriented participants to consider their skills development, learning, and growth. These outcomes stemmed from features paralleling those of the CorpoCapitalist prototype: rankings within the work community and feedback via popup windows after each evaluation and following a nine-article set (see Figures~\ref{fig:UI-differences} and~\ref{fig:UI-summary-differences}):

\begin{quote}
	This thing that it mentioned, your colleagues with whom you can compare those results, then I think that was like a door was opened -- like `hey, here is a chance to discuss this work', about evaluation and maybe that these are open problems. No right or wrong answers. It's just that this is how your colleagues evaluated the same cases. It kind of emphasised, I don't know, that I can discuss these tasks. And that there is a desire to support your employees and improve this system. (P3)
\end{quote}
\begin{quote}
	Yeah, [\ldots] I feel it reminds me of sitting exams in primary school and it's like... [\ldots] I like the feeling. I saw the AI results as like I'm doing my own homework. Then I compare to my classmates' homework, but because I've already done my own part, I have my own thoughts. So even [if] a classmate's result is different, I can still keep my thoughts. (P4)
\end{quote}

Moreover, users of this prototype experienced a greater sense of control over their work also because they could select news articles from the left sidebar in any order. The perceived difference in the level of control between the prototypes was also evident in the already above-presented statistical comparisons between the coping strategies in the two futures: again, while both futures held opportunities rather than threats, the participants found that the MyData future granted them more control over their situation, leaving room to get the most from the opportunities.

In sum, the prototypes evoked clearly different experiences of the two possible futures that we staged. Our discussion of these and the reasons for the divergence paves the way for examining the benefits that emerge from probing several possible futures instead of only one.

\subsection{Benefits from Triangulation on Possible Futures}

\begin{figure*}
  \includegraphics[width=\textwidth]{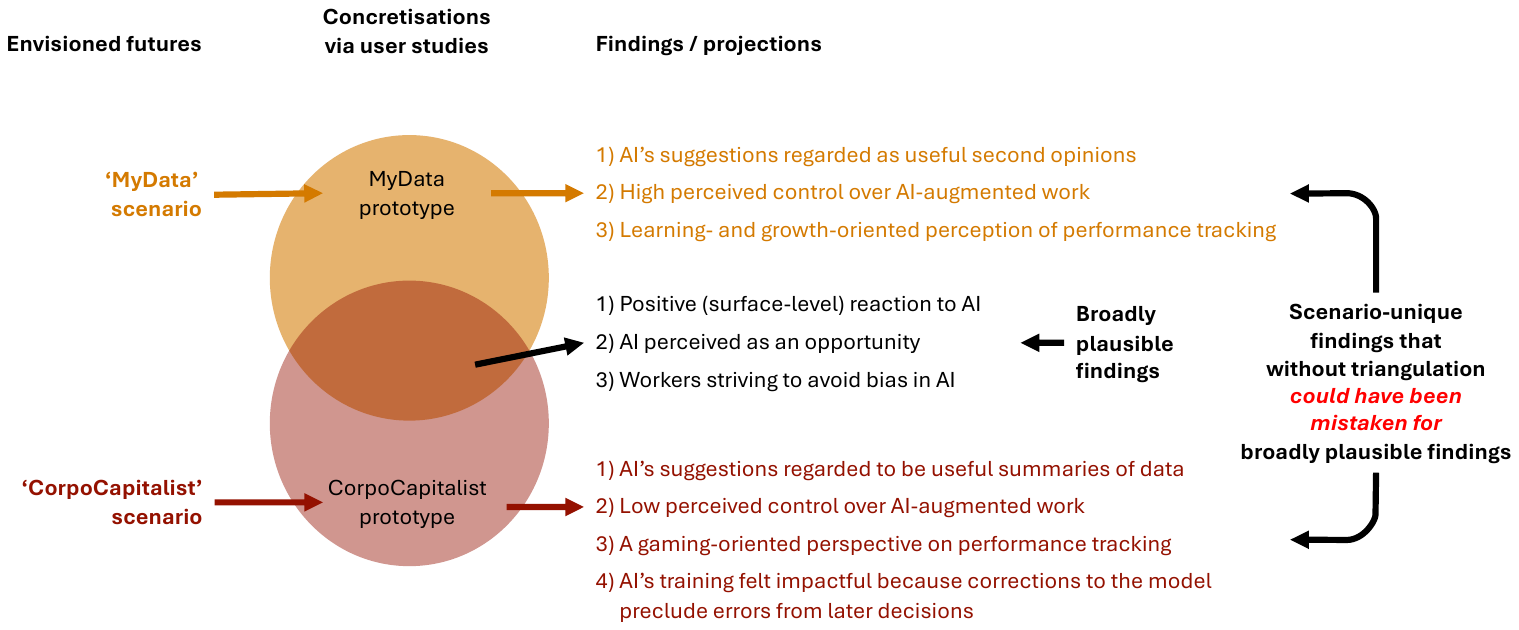}
  \caption{A demonstration of triangulation's value: our findings from triangulation across two possible futures, sorted by the future(s) in which the relevant pattern was observed.}
  \label{fig:findings-summary}
  \Description{Venn diagram about the findings of the study. The two circles of the diagram are MyData future and CorpoCapitalist future. Their overlapping findings are called "widely plausible" findings and were found from both futures. MyData future's and CorpoCapitalist future's unique findings are called "scenario-unique" findings. The figure lists following three widely plausible findings: 1. Positive (surface-level) reaction to AI, 2. AI perceived as an opportunity, 3. Workers strive to avoid bias from AI. The list of MyData's three scenario-unique findings are 1. AI’s suggestions are considered as useful second opinions, 2. High perceived control over AI-augmented work, 3. Learning and growth oriented perception to performance tracking. The list of CorpoCapitalist's four scenario-unique findings are 1. AI’s suggestions are considered as useful summaries of data, 2. Low perceived control over AI-augmented work, 3. Gaming-oriented perspective to performance tracking, 4. AI’s training feels impactful because corrections to the model remove errors from later decisions.}
\end{figure*}

Armed with valuable evidence demonstrating participants' interactions and experiences from decidedly different possible futures, 
we are now ready to reflect on how triangulation benefits studies on possible futures.
The discussion refers to Figure~\ref{fig:findings-summary}, a summarising visualisation that groups the empirical findings into three sets.

Where the circles in the diagram overlap, the two futures -- MyData and CorpoCapitalist -- share results in common. These findings may have \emph{broad plausibility}. We might, then, expect 1) generally positive reactions to the use of AI in news-article analysis, 2) workers in such settings perceiving AI as an opportunity rather than a threat, and 3) users developing varied techniques to avoid bias in their analysis and decision-making 
(the CorpoCapitalist condition displayed four distinct patterns in this regard, and the MyData prototype, irrespective of its lower risk of bias, still produced at least one). 
Because these findings demonstrate commonality between 
very different futures, it can be argued that their plausibility may hold across several possible futures, even ones not covered by the research space.

The non-overlapping portion at the top of the figure presents \emph{scenario-unique} results from participants' interactions and experiences in the MyData future. In that future only, 1) users found AI's suggestions particularly useful as second opinions after they had taken their own decision; 2) they saw AI as an opportunity and, additionally, had perceived control over its use; and 3) they associated the prototype's performance-tracking elements with avenues for learning, skills development, and growth.

Finally, the bottom of the figure lists another set of \emph{scenario-unique} findings, this time from the interactions and experiences in the CorpoCapitalist future. The findings exhibited several key divergences from the MyData future: 1) Because the suggestions were presented upfront, without regard for the user's level of desire to see them, they were considered differently. Users found value in them as data summaries (i.e., condensed versions of the articles), not straightforwardly as trustworthy recommendations awaiting confirmation as the final evaluation. 2) Though this future still struck participants as an opportunity, they perceived significantly lower control over the AI than in the MyData future. 3) Also, they approached the system's performance tracking (i.e., its assessment of their contribution to AI-model training) as a game that offered incentives to increase their ranking. 4) Finally, there was a further motivating factor alongside this, related to larger impacts of their corrections, through the AI's later operation in decision-making situations of a similar kind. 

The overlap-and-uniqueness-based categorisation of findings above concretises at least two benefits that triangulation can bring to studies of possible futures in HCI. We discuss these next. 

\subsubsection{Cross-Checking of Findings}

The first benefit stems from possibilities for \emph{cross-checking} \cite{fusch2018denzins,flick2018triangulation}. Its foundation, which was also called `cross-validation' in \nameref{sec:background}, builds on the distinction between scenario-unique and more broadly plausible findings, as crystallised on the far right in Figure~\ref{fig:findings-summary}. Had we studied only one of the two possible futures, we would not know which of the findings are unique to it and, hence, probably of limited plausibility. For example, a study confined to the MyData future could only have yielded a flat list of findings intermingling the uppermost and middle list in the figure. 

Left unaware of the probably broader plausibility of the latter set, we could equally well have identified the MyData-specific finding about learning-oriented experiences of performance-tracking as a significant result, perhaps even the main finding of our study. Mistaking that for a significant pattern that would extend across a wide range of possible AI futures, could have risked over-generalising our findings. At the same time, we would have ended up downplaying the importance of patterns listed in the middle -- patterns that, thanks to triangulation, we know we can deem more plausible. Also, we would have remained ignorant of that error.

\subsubsection{Deeper Insight through the Findings}

The content of Figure~\ref{fig:findings-summary} also anchors our discussion of the second benefit: alongside cross-checking, we gain opportunities for \emph{deeper insight} \cite{fusch2018denzins,flick2018triangulation}, in this case related most prominently to the deviations between the two possible futures studied.

Comparisons of findings between the MyData and CorpoCapitalist futures demonstrate one path to cultivating depth of insight. Overall, the MyData findings suggest that participants regarded the work with AI as motivating. A closer look at the reasons points to their freedom to evaluate news articles in any order; the availability of note-taking during their analyses, to enhance their skills; and the performance-tracking geared to informing about personal development. Interaction was personally motivating, and the participants engaged from a growth-oriented angle. The findings from the CorpoCapitalist future, in contrast, imply that its use was not as motivating personally, although many participants enjoyed its game-like user experience. In the absence of personal motivation factors, they justified its use in terms of potentially widespread positive impacts of their work as AI trainers and contributors to models.

As is befitting of rich insight, the analysis offers opportunities for more theorising. When taken together, the findings suggest the following: when the AI-augmented work was personally rewarding, participants tied their motivations to self-improvement and other positive intrinsic factors. In less personally rewarding work, in contrast, they justified their involvement via altruistic-seeming extrinsic and impersonal elements. If this interpretation is correct, it indicates that \emph{people's behaviour patterns and attitudes responding to AI-augmented knowledge work are dichotomous and might be opened up fruitfully in light of social psychology's rich body of work on motivation theories}. That literature offers theories of cognitive dissonance \cite{mcgrath2017dealing} and self-perception \cite{bem1967self-perception}, which have stood as competing models for explaining how people internally rationalise carrying out tasks that seem non-motivating. Theories of self-regulation \cite{hennecke2019doing} supply yet another lens for investigating the complex patterns to knowledge workers' use of AI. 

This investigation illustrates how triangulation can illuminate promising directions. Our work both has afforded deeper understanding -- of AI-augmented knowledge work's future and of two discrete types of motivation for it. 

\section{Discussion}
\label{sec:discussion}

Triangulation is an oft-recommended element of research in the social sciences and psychology especially, and it features in quantitative and qualitative methods alike, as the review in \nameref{sec:background} 
attests. 
While it has come to permeate much of HCI's user-research methodology landscape (e.g., see \cite{pettersson2018bermuda}), its full potential has yet to be exploited in our field. Given that HCI scholarship innovates and speculates about possible future technologies and their uses, we have attempted to cast light on this vast opportunity via the findings and suggestions reported here.

Namely, we point to triangulation on possible futures as a way of giving researchers better understanding of the plausibility of their results: do the findings apply uniquely in only one or another future, or more broadly (cf. Figure~\ref{fig:findings-summary}).
With our approach to studying multiple possible futures, not just one, we showed how staging two or more possible, suitably complementary futures and carrying out user studies in each of them equips researchers with tools to identify mutually distinct classes of findings whereby they can both cross-check the empirical findings and obtain deeper insight into futures.

Below, we extend our reflections pertaining to this fertile ground. Firstly, we look at several other mechanisms by which futures can be triangulated, beyond the approach demonstrated in the paper. We then consider the range of transformations that triangulation on possible futures could usher in for HCI research practice, in combination with what is required for actualising them.

\subsection{Types of Triangulation on Possible Futures}

Let us expand on the possibilities for triangulating with possible futures, in light of our experiences from the study and from contemplating it. We propose three types of triangulation for future-oriented studies in HCI, in a vein similar to the classic typology of triangulation by \citet{denzin1970research}. This facilitates deeper anchoring of the starting points for triangulation presented at the beginning of the paper (in Figure~\ref{fig:money-shot}).

\begin{figure*}
  \includegraphics[width=\textwidth]{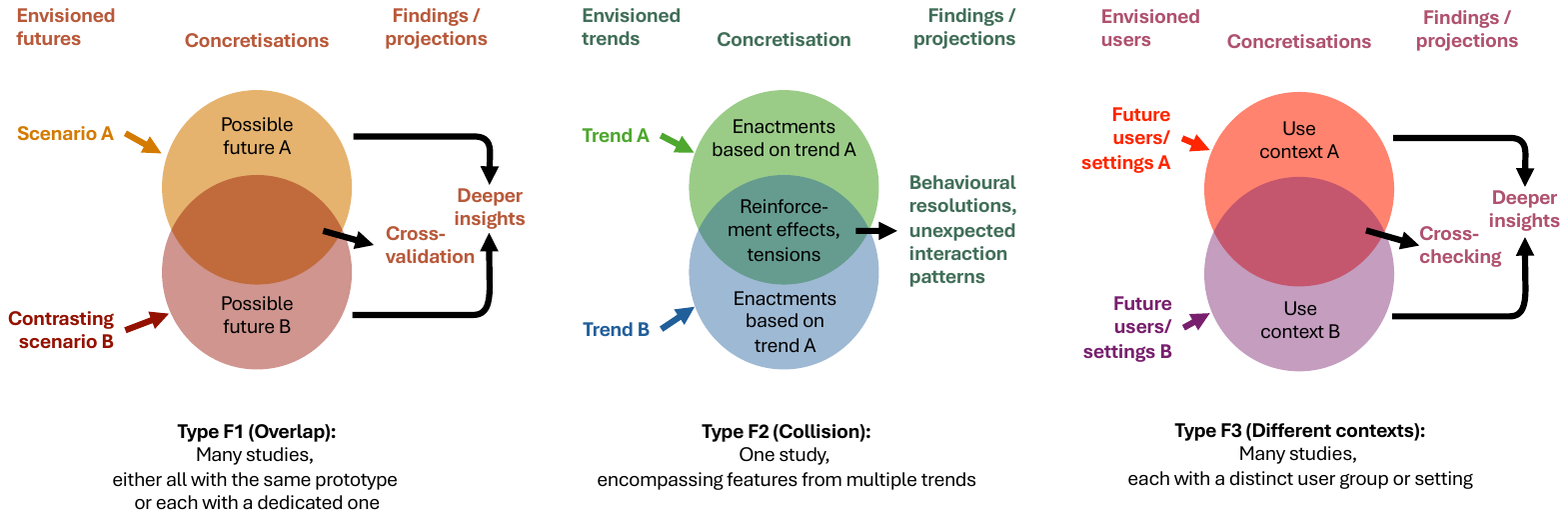}
  \caption{Three types of triangulation across possible futures.}
  \label{fig:triangulation-variants}
  \Description{Three Venn diagrams. Type F1 diagram shows triangulation of possible futures using contrasting futures, as in the present paper. Type F2 diagram shows triangulation that uses contrasting trends, where trends may reinforce each other, leading to new findings from their intersection. Type F3 diagram shows triangulation that applies different future user groups or settings and has many studies, each with a different user group or setting.}
\end{figure*}

Figure~\ref{fig:triangulation-variants} identifies these three types of future-oriented triangulation (we employ the prefix `F' to distinguish them from the original types introduced by \citet{denzin1970research}). \emph{Type F1} (Overlap) represents the class of research exemplified by our empirical study. Here, the researchers enact at least two futures, each one clearly distinct and also with a distinct user study (or arm thereof, as in our within-subjects research design). This may involve multiple prototypes, but having variants of a single design might simplify enactment. Our study is a case in point, supported with an overarching scenario of AI-augmented freelance knowledge work; though we employed different prototypes, Type F1 leaves room for studies in which only one prototype is necessary since scenarios or instructions might prove adequate for expressing the difference between/among the futures.

\emph{Type F2} (Collision) brings in a different futuring process. The triangulation here addresses future trends, weak signals, drivers, or critical uncertainties \cite{schwartz1991art} and any dependency relations among these. At pragmatic level, the fundamental difference from Type F1 is that Type F2 employs a single empirical study and prototype: elements from two trends are enacted in a single possible future. The benefit from this combination is that the future may be less naïve 
and more holistic \cite{moesgen2023designing} than in Type F1 counterparts, especially if the trends are chosen for their possibly conflicting implications. A user study of such a dynamic future may reveal user behaviours that would not ensue if only one or the other trend had been enacted. For example, such triangulation could be implemented for studying a possible future where companies' increasingly pervasive gathering of digital trace data would collide with the technological enablers that enrich citizens' opportunities to forge several digital identities opportunistically (e.g., digital wallets, digital ledger systems such as blockchain, and password managers).

Finally, \emph{Type F3} (Different contexts), which is visually similar in structure to Type F1, involves enacting contextual differences for the futures: in user groups, activities, locations, or other factors where the technology is used. The purpose is to study how appropriation of a technology may differ between conditions. The research design for such triangulation demands two or more user studies but may not require multiple prototypes.

Critical reflection on these types elucidates that greater effort needs to be expended for triangulating possible futures than for a typical single-future, single-prototype study (cf. Figure~\ref{fig:money-shot}). That said, overall, the cost of creating two variants of a prototype ought to be much lower than that of creating two entirely different prototypes. Such economy of scale should influence other staging activities \cite{salovaara2017evaluation} too, such as propping of a setting and preparation of mockup material. What this means is that expanding a single-future study to benefit from triangulation might not be as costly as one might presume. At the same time, putting in the effort confers potential for the study to yield more broadly plausible results 
and deeper insight, while it simultaneously encourages researchers to conceptualise their envisioned futures more thoughtfully. Such reflection can already increase a study's future-savviness.

We wish to emphasise that Types F1--F3 can function fruitfully in combination with \citet{denzin1970research}'s original types. For instance, our study employed methodological triangulation (his Type 4; Table~\ref{tab:denzin-triangulations})
by instrumenting the setting with interactive tasks, questionnaires, and interviews. 
A rich landscape of research designs is open to HCI from applying traditional triangulation (Types 1--4) in addition to future-oriented triangulation. A study might apply methodological triangulation by enacting a future in several ways -- for example, by arranging a lab study, an in-the-wild field study, and a VR-based field study \cite{makela2020virtual,simeone2022immersive} -- and then comparing the results so as to offset the weaknesses of each research method.

\subsection{Limitations and Future Work}

While the sample size in our exploratory study did not permit developing many interpretations from the quantitative responses, the paper attests to quantitative data's power to inform triangulation over possible futures, even if the methodological orientation is constructivist and qualitative material is a more typical source. We posit that quantitative analyses' results can serve as a basis for interpreting possible futures. 

Reflecting on our statistical analyses led us to consider more profound questions about the value of such analysis in future-oriented HCI studies. The principle that statistical analyses should be hypothesis-based might conflict with the epistemology of studying possible futures. That is, hypotheses are truth statements, but the truthfulness of future states is not testable. Nevertheless, it remains important to seek solid criteria for credibility in the field's future-oriented studies. Our paper's epistemology built on the ideas of objective shared reality and social construction of understandings of that reality, which included also representations of that reality's futures \cite{clardy2022what}. By adopting a similar vantage point, scholars can carry out studies that present counter-evidence to earlier (triangulated) findings, akin to how positivist researchers might strive for replication and falsification (though with the caveat that contrary evidence does not constitute refutation, on account of future representation's constructivist nature). But this epistemological position requires critical scrutiny that was beyond this paper's scope. A philosophical analysis on the epistemology on possible futures research would deserve its own paper.

Before moving on, however, we ought to say something about practical implications of the constructivist nature of triangulating over possible futures. One might wonder how many futures a study's triangulation should encompass and what perspectives ought to guide the choice of those futures. A constructivist response would be that there are many valid answers, though their credibility levels may differ. A researcher with deep, broad-based knowledge of multiple theories, areas of research, etc. is equipped for insight-affording viewpoints and solid justifications -- thus with better-grounded justifications for the types of triangulation and their adequate number. 

Therefore, our proposal presupposes the researcher's prior awareness of meaningful contrasts on which to base the triangulation.  Above, we suggested three possible designs (F1--F3) and recommended that the differences are grounded theoretically or conceptually.
If triangulation on possible futures becomes commonplace among HCI scholars, the experience of the value accumulated should aid in developing principles for their research designs. This can help avoid new methodological pitfalls in the method, such as that the overlapping findings from the two studies may sometimes not yield broadly plausible information, but only point out irrelevant commonalities.

\section{Conclusion}

In the title of a UI-testing paper published in \citeyear{tohidi2006getting}, \citet{tohidi2006getting} stated that `testing many is better than one'. The paper's main argument was that testing multiple alternative designs enables more accurate comparative evaluation. Although the article was about UIs rather than possible futures and did not explicitly discuss triangulation, it served as a source of inspiration for the study reported upon here -- so much so that we decided to emulate its title in our own. 

Similarly to \citeauthor{tohidi2006getting}, we recognised significant value in arranging multi-pronged studies of a given phenomenon. Therefore, we articulated the benefits via a conceptual review, then conducted a user study that operationalised our conceptual-level proposal, finally extending our examination of triangulation of possible futures to propose three new, distinct types of future-oriented triangulation. The nascent stage of this methodological approach notwithstanding, our findings were promising, warranting further research. Given the novelty of our proposal, we would expect additional types of futures triangulation to be identifiable, and that further modelling could develop our proposed types further. This opens new, conceptually and methodologically grounded paths into an exciting but slippery research topic: the study and anticipation of possible futures, a pivotal concern for our future-oriented research field.

\begin{acks}
The authors thank Anton Poikolainen Rosén, whose mention of `triangulation of possible futures' served as a seed for much of the investigation presented here.
We are grateful also to Jesse Haapoja, for discussion of social-psychological theories on motivation, and for comments on draft versions from Anton Poikolainen Rosén, Camilo Sanchez, Tim Moesgen, and Kaisa Savolainen. This research was supported financially by the Research Council of Finland (grant 330124).
\end{acks}

\bibliographystyle{ACM-Reference-Format}
\bibliography{references}

\appendix

\section{The Post-Interaction Questionnaire and Interview}
\label{sec:post-scenario-interview}

The following data were collected after both the `morning' and the `afternoon' work session.

\subsection*{The Attitude Questionnaire}

1 = Strongly Disagree, 2 = Disagree, 3 = Somewhat Disagree, 4 = Am Neutral, 5 = Somewhat Agree, 6 = Agree, 7 = Strongly Agree 

\begin{tabular}{|p{4.3cm}|c|c|c|c|c|c|c|}
\hline
& 1 & 2 & 3 & 4 & 5 & 6 & 7\\
\hline
\textbf{Attitude to using the system} & & & & & & &\\
\hline
Using this system is a good idea & & & & & & &\\
\hline
The system makes work more interesting & & & & & & &\\
\hline
Working with the system is fun & & & & & & &\\
\hline
I like working with the system & & & & & & &\\
\hline
\textbf{Compatibility with one's preferred work method}  & & & & & & &\\
\hline
Using the system is compatible with all aspects of my work & & & & & & &\\
\hline
I think that using the system fits well with the way 1 like to work & & & & & & &\\ 
\hline
Using the system fits my work style & & & & & & &\\ 
\hline
\textbf{The task's perceived significance} & & & & & & &\\
\hline
The results of my work are likely to affect the lives of other people significantly & & & & & & &\\
\hline
The job itself is very significant and important in the broader scheme of things & & & & & & &\\
\hline
The job has a large impact on people outside the organisation & & & & & & &\\
\hline
\end{tabular}

\subsection*{Questions about Coping}

\begin{itemize}
    \item Do you see this kind of work arrangement as a threat or as an opportunity, and why?
    \item If the organisation were to offer such a work arrangement, what kind of control would you need so that it would work well for you? High control or low control? Why, and in what way could this be done?  
    \item Would you describe your attitude to this work arrangement as most closely resembling benefits maximising, benefits satisficing, disturbance handling, or self-preservation? 
\end{itemize}

The following definitions were provided alongside the question above:

\begin{itemize}
\item \emph{Benefits maximising} = `You have a chance to benefit from the arrangement and have the capacity for doing so.'
\item \emph{Benefits satisficing} = `You have a chance to make use of the situation, but you can’t do much about it – things just happen.'
\item \emph{Disturbance handling} = `The arrangement is not beneficial to you, but you have capacity to minimise its impact.'
\item \emph{Self-preservation} = `The arrangement is not beneficial for you, and you are mostly at its mercy.'
\end{itemize}

\section{The Concluding Interview}
\label{sec:concluding-interview}

These questions were asked after both of the scenarios connected with possible futures had been studied. Printouts of screenshots from the user interface of both prototypes (CorpoCapitalist and MyData) were available, to support memory.

\subsection*{Differences}

\begin{enumerate}
    \item What differences did you notice in how you used the system in the morning vs. the afternoon? Were there any patterns that you followed with one particular system, or with both? 
    \item How satisfied were you with your ability to perform your tasks in the morning and in the afternoon? Did you feel that the AI system affected your performance? If so, how? 
    \item If using these types of systems were to become a common way of working, would you perceive them as a helpful tool or as a potential threat to your role? 
    \item How did it feel to carry out these two jobs? 
    \item How would you describe the work for Invezterr [the MyData future]? How would you describe the work for Stockify [the CorpoCapitalist future]?  
\end{enumerate}

\subsection{The Jagged Frontier}

\begin{enumerate}
    \item How much did you feel able to rely on the AI in each work setting? Did you consider the AI's help generally reliable? 
    \item For each prototype: Did you find cases in which, in your opinion, the AI was outright wrong? Please identify the case and how the AI was wrong, if you can recall this.
    \item Which prototype do you think involved more effort and engagement in completing your tasks? Please explain why it feels that way. 
\end{enumerate}

\subsection*{Learning in Knowledge Work}

\begin{enumerate}
    \item Which work do you think better supported your development as a professional? Why? 
    \item How did you feel about the feedback that the systems provided?
\end{enumerate}
 
\subsection*{Reflective Questions about STEEPLE Categories}

Could you elaborate on the positive, negative, and other kinds of opinions you have about the ways you worked with the two systems, and on the work arrangement overall?

\begin{itemize}
\item From a \emph{social} point of view
\item From a \emph{technological} standpoint
\item From an \emph{economic} point of view
\item From an \emph{ethics} angle
\item From a \emph{political} point of view
\item From a \emph{legal} point of view
\item From an \emph{ecological} point of view
\end{itemize}

\end{document}